\documentclass[runningheads]{llncs}
\usepackage[htt]{hyphenat}
\usepackage{amssymb}
\usepackage{dirtytalk}
\usepackage{csquotes}
\usepackage{listings}
\usepackage{footmisc}
\usepackage{footnote}
\usepackage[dvipsnames]{xcolor}
\usepackage{graphicx}
\definecolor{codegreen}{rgb}{0,0.6,0}
\definecolor{codegray}{rgb}{0.5,0.5,0.5}
\definecolor{codepurple}{rgb}{0.58,0,0.82}
\definecolor{backcolour}{rgb}{0.96,0.96,0.96}

\lstdefinestyle{cstyle}{
    language=C,
    frame=lines,
    backgroundcolor=\color{backcolour},   
    commentstyle=\color{codegreen},
    keywordstyle=\color{codepurple},
    numberstyle=\tiny\color{codegray},
    stringstyle=\color{magenta},
    basicstyle=\ttfamily\bfseries\footnotesize,
    breakatwhitespace=false,         
    breaklines=true,                 
    captionpos=b,                    
    keepspaces=true,                 
    numbers=left,                    
    numbersep=5pt,                  
    showspaces=false,                
    showstringspaces=false,
    showtabs=false,                  
    tabsize=2
}

\lstdefinestyle{pystyle}{
    language=Python,
    frame=lines,
    backgroundcolor=\color{backcolour},   
    commentstyle=\color{codegreen},
    keywordstyle=\color{codepurple},
    numberstyle=\tiny\color{codegray},
    stringstyle=\color{magenta},
    basicstyle=\ttfamily\bfseries\footnotesize,
    breakatwhitespace=false,         
    breaklines=true,                 
    captionpos=b,                    
    keepspaces=true,                 
    numbers=left,                    
    numbersep=5pt,                  
    showspaces=false,                
    showstringspaces=false,
    showtabs=false,                  
    tabsize=2
}

\makesavenoteenv{tabular}
\makesavenoteenv{table}

\begin{document}
\title{Performance of the Vipera framework for DSLs on micro-core architectures}
%
%
\author{Maurice Jamieson\inst{} \and
Nick Brown\inst{}}
\authorrunning{M. Jamieson and N. Brown}
%
\institute{{EPCC, The University of Edinburgh\\
Edinburgh, UK\\
\email{maurice.jamieson@ed.ac.uk}\\
\url{https://www.epcc.ed.ac.uk}}}
\maketitle              
\begin{abstract}
Vipera provides a compiler and runtime framework for implementing dynamic Domain-Specific Languages on micro-core architectures. The performance and code size of the generated code is critical on these architectures. In this paper we present the results of our investigations into the efficiency of Vipera in terms of code performance and size.

\end{abstract}
\section{Introduction}
 In order to reduce the power consumption of new High-Performance Computing (HPC) machines, the use of hybrid HPC architectures with graphics processing units (GPUs) as accelerators has increased, such as the 4:1 ratio of GPUs to central processing units (CPUs) per node of the new OLCF Frontier exascale supercomputer\cite{noauthor_frontier_2021}. Other novel architectures for HPC have been introduced, including innovative \emph{micro-core}\footnote{Although the term \emph{manycore} is commonly used, we define micro-cores as manycores with extremely small amounts of on-chip, scratchpad RAM (circa 32 - 64KB) without hardware cache support.} processor architectures that consist of many, low energy cores combined with small amounts of memory on a single chip, such as the 256 core Kalray MPPA
 , the 256 core Sunway SW26010
 , the 1024 Adapteva Epiphany-V
 and the 2048 core PEZY-SC2
 . These micro-core architectures have the promise of overcoming the power wall due to the high energy efficiency of their designs, for example, the class-leading 70 GFLOPS per Watt of the the 64-core Adapteva Epiphany-IV 
 \cite{green500_june_2021}. Whilst these architectures provide the high energy efficiency and low overall power consumption levels, micro-cores are notoriously difficult to program and take advantage of; each technology is different with its own idiosyncrasies, such as the topology of the Network-on-Chip (NOC), and they each present a different low-level interface to the programmer. Although manufacturers have made great progress in developing the hardware, parallel programming and compilation techniques have not evolved quickly enough to exploit this effectively\cite{kudlur_orchestrating_2008}. Fundamentally, writing parallel, scalable code is difficult and requires the programmer to consider multiple levels of parallelism to get good performance\cite{lyberis_myrmics_2012}. 
However, to date, these technologies have tended to result in significant performance overheads, required the programmer to ensure their code fits within the limited on-chip memory, provided limited choices around data location and size, and provided little, if any, portability across architectures. As evidenced by ePython\cite{brown_epython_2016}, a Python interpreter for the Epiphany-III, dynamic programming languages can significantly reduce the programming effort required to overcome these complexities in comparison to the provided, low-level C software development kits (SDKs)\cite{jamieson_brown_liu_epython_2020}.
 
 In this paper we present the investigations into the efficiency of our Vipera framework for dynamic programming languages, in terms of code performance and size, relative to handwritten (native) C, on a variety of micro-core and traditional CPU architectures.
 
\section{Background and related work}
Whilst Python is currently the most popular programming language\cite{tiobe}, its use of an interpreter results in performance significantly slower than statically compiled languages, such as C and Fortran. This has driven the need to overcome the performance overhead of the interpreter and the restrictions imposed by the global interpreter lock (GIL). 
This has resulted in technologies to increase the performance of existing Python codes through the compilation to native code, including Cython\cite{behnel_cython_2011}, MicroPython\cite{noauthor_micropython_nodate}, Numba\cite{lam_numba_2015}, Copperhead\cite{catanzaro_copperhead_2011}, Parakeet\cite{rubinsteyn_parakeet_2012}, ALPyNA\cite{jacob_python_2019} and PyCUDA\cite{noauthor_pycuda_nodate}. The high-level approach of Numba, Copperhead and Parakeet is similar, whereby they define an embedded domain specific language (eDSL) and utilise Python \emph{function decorators} (directives) to annotate the code to be compiled to native code or offloaded to GPUs. ALPyNA adopts a different technique to generating GPU code than the eDSL and function decorator approach. Rather than requiring the programmer to select and annotate the Python functions that will be generated as GPU kernels, ALPyNA analyses loop data dependencies and performs automatic loop parallelisation to generate CUDA kernels for GPUs. However, unlike Numba, Copperhead, Parakeet and ALPyNA, PyCUDA does not abstract the generation of GPU code but instead embeds CUDA C code directly within the Python source code. 
MicroPython performs the compilation of bytecode to native code on the device\cite{noauthor_micropython_update_2013} similar to JIT except that the bytecode is not profiled as is common for JIT compilers, rather the bytecode is just lowered to native code. An alternative approach was taken for Vipera, similar to that employed by the Pallene / Titan compiler\cite{gualandi_pallene_2018} for Lua\cite{ierusalimschy_lua_implementation_nodate}. Here, the source language compiler, running on the host, emits C source code that is then compiled to generate native binary executables.

\subsection{Vipera dynamic language framework}
The Vipera\cite{brown_vipera_2022} framework was created to support the development of dynamic languages on micro-core architectures. The framework consists of a layered architecture with components running on the host and micro-core devices. Vipera manages the compilation of code, the transfer and launch of kernels on the micro-core devices, and the transfer of data. vPython is a development of ePython, a subset of the Python programming language specifically designed for micro-core architectures. Vipera provides two implementations of this; the first compiles down to bytecode that executes on a tiny virtual machine (c. 24KB on the Adapteva Epiphany-III\cite{brown_epython_2016}) running on the device and the second generates Olympus abstract machine code that is compiled to provide device native code. In this paper we will focus on the Olympus abstract machine version of vPython.

vPython can either be run \emph{standalone} on the device or as a Domain-Specific Language (DSL) within Python running on the host, offloading kernels for execution to the device. More information on the parallel programming, offloading and dynamic code loading capabilities of the language can be found in \cite{jamieson_brown_liu_epython_2020} and \cite{jamieson_compact_2021}.


\section{Benchmarking}
\subsection{CPU selection}\label{sec:cpuselect}
In order to support the assessment of the Vipera vPython compiler and Olympus abstract machine, a number of different platforms and processors were selected, including the Adapteva Epiphany-III
, Xilinx MicroBlaze
and PicoRV32
RISC-V micro-cores and the AMD64 (x64), ARM Cortex-A9 (ARM32), MIPS32, SPARCv9 and U740 RISC-V (RISCV64) traditional CPUs. As processor ISAs can have a significant impact on both the compiled kernel performance and binary size, the CPUs were selected to test the impact of the Olympus abstract machine design and to test the portability of Olympus between 32 bit and 64 bit processors with varying alignment constraints and byte ordering. 

\ifdefined\CPUTABLE
\begin{table}[ht!]
\caption{Test environment board specifications}
\centering
\footnotesize
\begin{tabular}{|lllll|}
\hline
\textbf{Machine}  & \textbf{Arch} & \textbf{CPU} & \textbf{RAM} & \textbf{OS}\\ \hline
\textbf{Adapteva Parallella} & {32 bit} & 2x\footnote{Refers to the number of cores within the CPU.} 650MHz Cortex-A9 (ARM) & 1GB & Ubuntu 16\\
\textbf{Adapteva Parallella} & {32 bit} & {16x 600MHz Epiphany-III} & 32KB*\footnote[1]{* Memory per core.} & bare metal\\
\textbf{Xilinx PYNQ-Z2} & {32 bit} &  {8x 100MHz MicroBlaze} & {64KB*} & {bare metal} \\
\textbf{Xilinx PYNQ-Z2} & {32 bit} & {8x 100MHz PicoRV32 RISC-V} & {64KB*} & {bare metal} \\
\textbf{Creator Ci20} & {32 bit} & {2x 1.2GHz XBurst MIPS32}  & 1GB & Debian 10  \\
\textbf{HP 15-g093} & {64 bit} & {4x AMD A4-6210 AMD64}  &  4GB & {WSL}\footnote{Windows Subsystem for Linux on Microsoft Windows 10.}\\
\textbf{Sun ULTRA 5} & {64 bit}\footnote{Olympus abstract machine and native C benchmark codes compiled to 32 bit binaries, with 32 bit integers and single-precision floating point on the SPARCv9.} & {1x 400MHz UltraSPARC-IIi} &  512MB & {Solaris 10} \\
\textbf{HiFive Unmatched} & {64 bit} & {4x 1GHz U740 RISC-V} &  16GB & {Ubuntu 21} \\ \hline
\end{tabular}
\label{tbl:testmachines}
\end{table}
\fi 


For the selected benchmarks, LINPACK\cite{dongarra_linpack_2003} and the Sieve of Eratosthenes\cite{gilbreath_high-level_1981}, the source vPython codes were compiled to Olympus abstract machine C source code and wrapped by Eithne\cite{noauthor_maurice_nodate} API calls for execution on a single core of the CPUs. 

\ifdefined\BENCHMARKOVERVIEW
\subsection{LINPACK overview}\label{sec:linpackbench}
The LINPACK benchmark\cite{dongarra_linpack_2003} measures the floating point performance of a computer by solving a matrix problem using LU decomposition. It is a long-established benchmark, having been introduced in 1979, and is the standard benchmark used to rank supercomputer performance for the Top500\cite{November2019TOP5002019} list. 
LU factorisation is commonly used in scientific and industrial applications, such as design automation, machine learning and signal processing\cite{jackson_lufactorisaton_2016}. The C version\cite{noauthor_linpack_bench_nodate} was selected to compare the performance (MFLOPS\footnote{Million Floating Point Operations per Second}) of the Olympus abstract machine\footnote{An vPython version of LINPACK was written and compiled to Olympus abstract machine code.} against native C, at GCC compiler optimisation levels \texttt{-Os} and \texttt{-O3}. For the Olympus abstract machine investigations, a serial version of the LINPACK code was run on a single core. Due to the extremely small memory available on the micro-core devices, the problem size $n$ was 50 and for traditional CPUs $n=1000$. 

\subsection{Sieve of Eratosthenes (Byte Sieve) overview}\label{sec:sievebench}
The Sieve of Eratosthenes benchmark\cite{gilbreath_high-level_1981}, often referred to as \emph{Byte Sieve}, is commonly used to test compiler code generation performance and efficiency\cite{keil_sieve_nodate}\cite{segger_sieve_nodate}. The standard C and a new vPython version of the benchmark were used to determine the performance of integer array access and looping constructs of the Olympus abstract machine relative to native C, to augment the LINPACK benchmark that performs floating point array access and calculations. The standard Sieve benchmark is serial and both versions were run on a single core of all the CPUs. Like LINPACK, due to the limited memory on the micro-core devices, the flag array size was reduced ($SIZE=4095$) on the Epiphany-III, MicroBlaze and PicoRV32 micro-cores, on the other CPUs, per the original benchmark, $SIZE=8190$. 
\fi

\subsection{LINPACK performance}\label{sec:codeperf}
\begin{figure}[h!]
	\centering
	\includegraphics[width=0.99\textwidth]{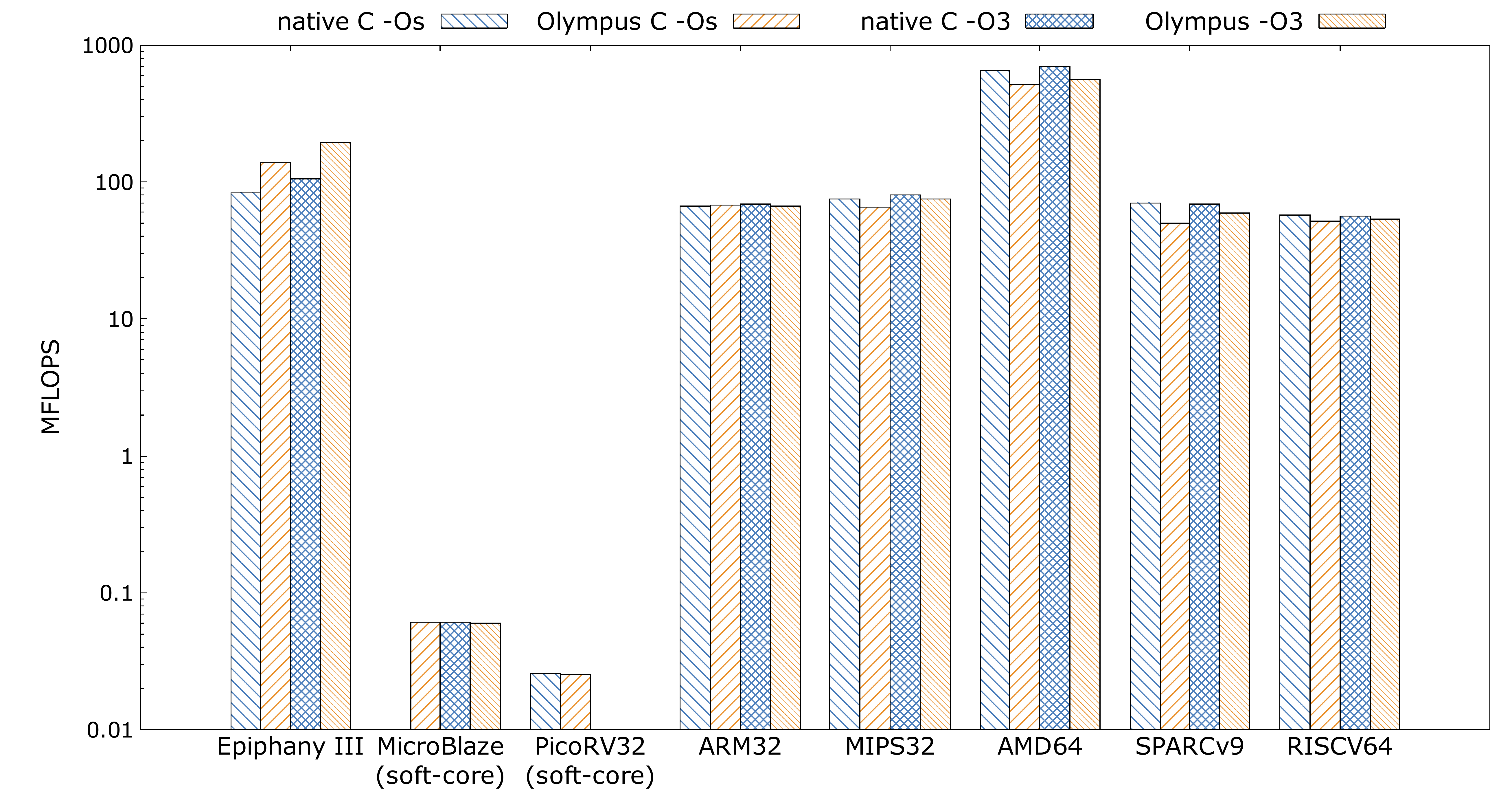}
	\caption{LINPACK benchmark native C and Olympus kernel performance (log scale)}
	\label{fig:linpackperformance}
\end{figure}
Figure \ref{fig:linpackperformance} shows the single-core performance results for LINPACK\footnote{Due to the extremely small memory available on the micro-core devices, the problem size $n$ was 50 and for traditional CPUs $n=1000$.} on the target processor architectures, compiled using the \texttt{-Os} and \texttt{-O3} compiler optimisation levels. Whilst the results vary widely across the architectures, the performance difference between the Olympus and native C kernels is very small. However, the Olympus LINPACK kernel compiled at \texttt{-Os} is 1.7 times faster than native C on the Epiphany-III and is marginally faster (1.5\%) on the ARM32. Although the performance advantage of Olympus kernels on the ARM32 is reversed at \texttt{-O3}, where native C is 4.6\% faster, the advantage is actually slightly increased at \texttt{-O3} on the Epiphany-III to 1.8 times faster than native C. On the other architectures, native C is between about 1.2\% on the MicroBlaze and 42\% on the SPARC faster than Olympus at \texttt{-Os} and between about 7.2\% on the MIPS32 and 24\% on the AMD64 faster at \texttt{-O3}.

\ifdefined\DATATABLES
\subsection{Tailoring for RISC-V}
\begin{table}[h!]
\caption{LINPACK kernel performance (MFLOPS)}
\centering
\footnotesize
\begin{tabular}{|lllll|}
\hline
\textbf{CPU}  & \textbf{Olympus (-Os)} & \textbf{Olympus (-O3)} & \textbf{native C (-Os)}  & \textbf{native C (-O3)} \\ \hline
\textbf{Epiphany-III}  & 139.36 & 193.53 & 83.99 & 105.99 \\
\textbf{MicroBlaze}  & 0.0604 & 0.0605 & 0.0611 & 0.0612  \\
\textbf{PicoRV32} & 0.0252 & N/A\footnote{On the PicoRV32 at GCC optimisation level \texttt{-O3}, both the Olympus and native C benchmarks abort with a \emph{\say{The matrix A is apparently singular}} LINPACK error. As both versions of the benchmark fail, it is suspected that there is a code generation issue for the PicoRV32 at this level of optimisation that requires further investigation.} & 0.0256 & N/A \\ 
\textbf{ARM32} & 67.74  & 66.61  & 66.71  & 69.67 \\
\textbf{MIPS32} & 65.90  &  75.60 & 75.40  & 81.077 \\
\textbf{AMD64}  & 520.98  &  563.03 & 652.46 &  695.57 \\
\textbf{SPARCv9} & 49.81  &  59.40 & 70.78 &  68.96 \\
\textbf{RISCV64} & 51.12  &  57.48 & 53.46 &  56.57 \\ \hline
\end{tabular}
\label{tbl:linpackperf}
\end{table}
\fi


Analysing the performance advantage of Olympus kernels over native C on the Epiphany-III and at \texttt{-Os} on the ARM32 requires knowledge of the peculiarities of the Epiphany-III and looking at the assembly language generated by the C compiler. In the case of the Epiphany-III, there are four modes for the floating point unit (FPU) that can be specified at compile time\cite{noauthor_epiphany_2013}.
\ifdefined\EPIPHANYMODES
\begin{itemize}
    \item \emph{caller}: Any mode at function entry is valid, and retained or restored when the function returns, and when it calls other functions.
This mode is useful for compiling libraries or other compilation units you might want to incorporate into different programs with different prevailing FPU modes, and the convenience of being able to use a single object file outweighs the size and speed overhead for any extra mode
switching that might be needed, compared with what would be needed with a more specific choice of prevailing FPU mode.
    \item \emph{truncate}: This is the mode used for floating point calculations with truncating (i.e. round towards zero) rounding mode. That
includes conversion from floating point to integer.
    \item \emph{round-nearest}: This is the mode used for floating point calculations with round-to-nearest-or-even rounding mode.
    \item \emph{int}: This is the mode used to perform integer calculations in the FPU, e.g. integer multiply, or integer multiply-and-accumulate.
\end{itemize}
\fi 
The default FPU mode is \emph{caller}, which results\footnote{The FPU mode comparisons were all performed using the \texttt{-O3} compiler optimisation level.} in native C kernels being 1.7 times faster than Olympus. The \emph{truncate} FPU mode does not provide a significant improvement (2.1\%) of native C kernels over Olympus. The \emph{round-nearest} mode provides a 2.1 times performance improvement of native C over the  Olympus abstract machine. The \emph{int} FPU mode, executing integer operations as well as floating point operations in the FPU, delivers a 1.66 and 1.83 times performance advantage of Olympus kernels over native C at for \texttt{-Os} and for \texttt{-O3}, respectively. This result is surprising but considering that the Epiphany-III is a superscalar design that can execute two floating point operations and one integer instruction per clock cycle\cite{E16G301datasheet}, it is possible to surmise that the Olympus \emph{mnemonics} can take advantage of the additional two integer operations per clock cycle afforded by the \emph{int} FPU mode and prevent the pipleline from stalling. 

The minor performance advantage of Olympus over native C on the ARM at the \texttt{-Os} compiler optimisation level can be explained by the additional 21 \texttt{APSR\_nzcv} opcodes in the native C kernel. This opcode transfers the floating-point status flags are transferred the ARM application program status register (APSR) and, as \cite{noauthor_neon_2013} state:
\begin{quote}
\textit{These instructions stall the ARM until all current NEON or VFP operations complete.}
\end{quote}
It is also interesting to determine from the disassembly listing of the ARM Olympus kernel that the ARM NEON vector / SIMD instructions (e.g. \texttt{VLDR}, \texttt{VLMUL} and \texttt{VSTR}) are being issued by the C compiler for the Olympus mnemonics, thereby taking advantage of this parallel processing capability of the ARM processor for the LINPACK benchmark.

\subsection{LINPACK code size}\label{sec:codesize}
\begin{figure}[h!]
	\centering
	\includegraphics[width=0.99\textwidth]{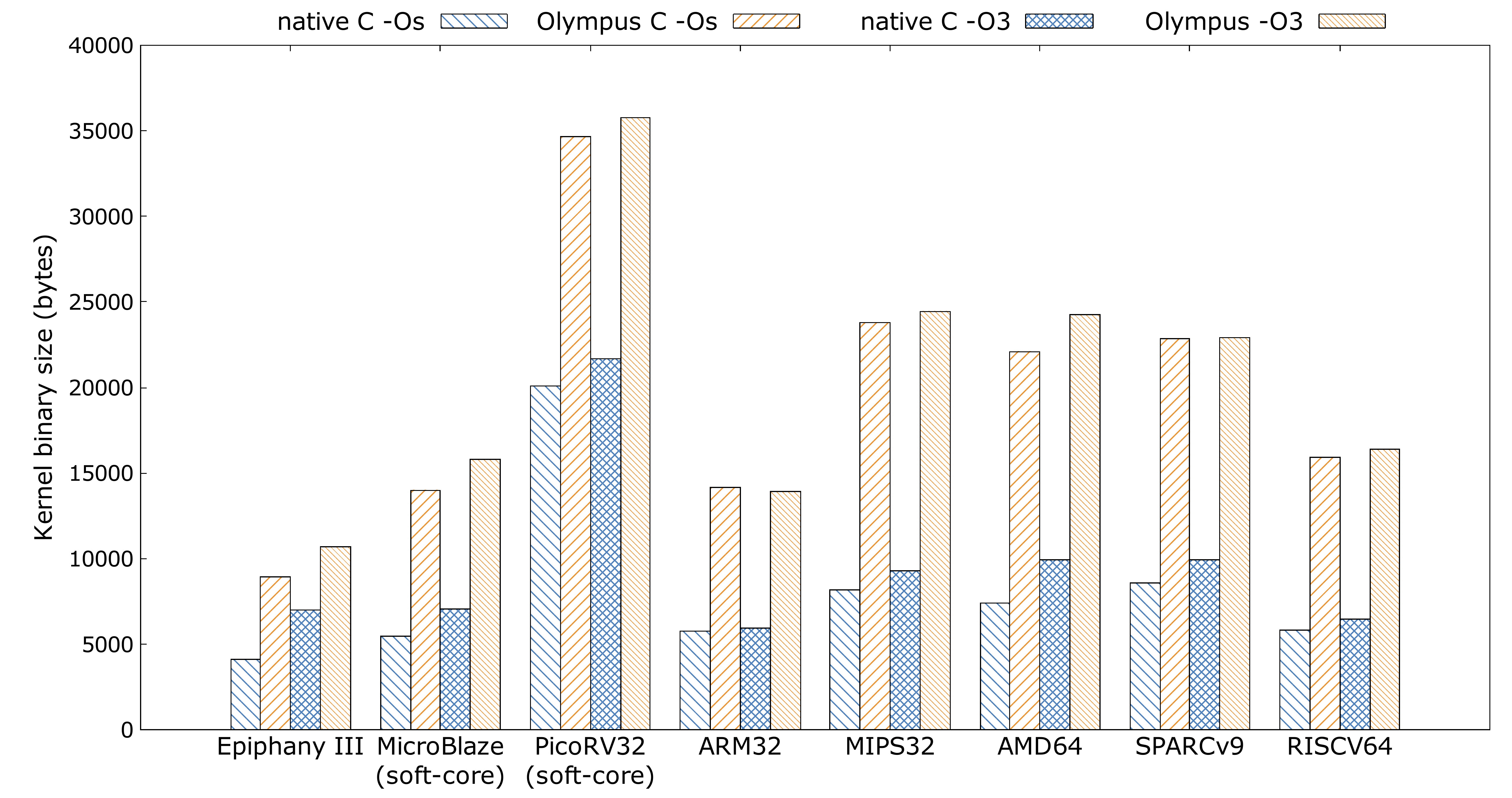}
	\caption{LINPACK benchmark native C and Olympus kernel size}
	\label{fig:linpacksize}
\end{figure}

 Figure \ref{fig:linpacksize} illustrates that the C kernels are significantly smaller than the Olympus kernels on all platforms, at GCC optimisation levels \texttt{-Os} and \texttt{-O3}, for the LINPACK benchmark. The difference in kernel size ranges from around 1.5 times bigger than native C on the Epiphany-III to 2.6 times bigger on the MIPS32, using \texttt{-O3}. Interestingly, the difference ranges from around 2 times bigger than native C on the Epiphany-III to around 3 times bigger on the MIPS32 and AMD64. This suggests that the Olympus mnemonics generate \emph{wordy} C code, whereby a significantly larger number of underlying operations (machine opcodes) are generated by the C compiler in comparison to the equivalent native C operation. 
 However, it should be noted that the Olympus kernels include a full compacting heap manager and other runtime functions required to support the dynamic features of ePython that are absent from the static native C LINPACK kernel. 
 
The figures for the MicroBlaze reflect the use of the floating-point emulation option for the LINPACK benchmark. Unsurprisingly, the code size difference is greater on the MicroBlaze in comparison to the Epiphany-III, at between 2 and 2.6 times larger (for both compiler optimisation levels), due to the increased number of operations generated by the Olympus mnemonics over native C, which is amplified by the floating-point emulation code required for the MicroBlaze LINPACK benchmark. 
\ifdefined\DATATABLES
\begin{table}[h!]
\caption{LINPACK kernel code size (bytes)}
\centering
\footnotesize
\begin{tabular}{|lllll|}
\hline
\textbf{CPU}  & \textbf{Olympus (-Os)} & \textbf{Olympus (-O3)} & \textbf{native C (-Os)}  & \textbf{native C (-O3)} \\ \hline
\textbf{Epiphany-III}  & 8904 & 10708 & 4124 & 7012 \\
\textbf{MicroBlaze}  & 14008 & 15780 & 5456  & 7048  \\
\textbf{PicoRV32} & 32484 & 33348 & 19056 & 20420 \\ 
\textbf{ARM32} & 14171  & 13935  & 5731  & 5909 \\
\textbf{MIPS32} & 23772  &  24460 & 8166  & 9258 \\
\textbf{AMD64}  & 22111  &  24279 & 7376 &  9929\\
\textbf{SPARCv9} & 22876  &  22887 & 8602 &  9917 \\ 
\textbf{RISCV64} & 15903  & 16415 & 5795 & 6445 \\ \hline
\end{tabular}
\label{tbl:linpacksize}
\end{table}
\fi 
There is up to a 20\%  advantage, on the Epiphany-III, in terms of code size in selecting \texttt{-Os} over \texttt{-O3}. However, for the SPARCv9 the advantage is minimal (0.048\%) and is actually detrimental on the ARM32 (-1.67\%). Overall, there is an average increase in code size of 7.5\% selecting \texttt{-O3} over \texttt{-Os}, which needs to be considered relative to any performance advantage gained by selecting the higher compiler optimisation level. For a micro-core architecture, such as the Epiphany-III, the code size saving of 20\% (approximately 1.8KB) could be significant. Therefore, it is important to understand any performance differences between the two compiler optimisation levels. 

\subsection{Sieve of Eratosthenes performance}\label{sec:sievesperformance}
\begin{figure}[h!]
	\centering
	\includegraphics[width=0.99\textwidth]{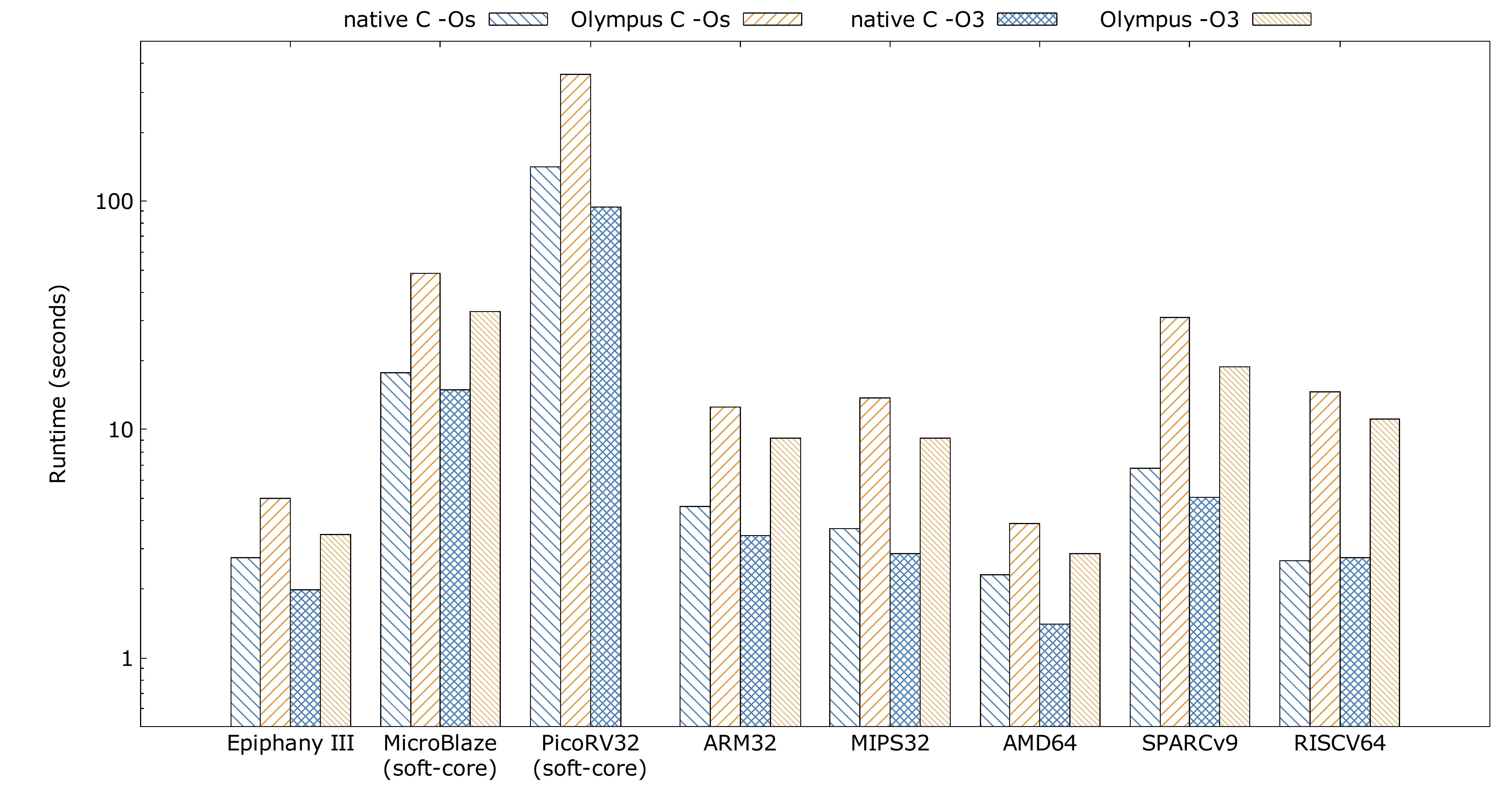}
	\caption{Sieve benchmark native C and Olympus kernel runtime (log scale)}
	\label{fig:sieveperformance}
\end{figure}

The LINPACK benchmark tests the floating point performance of the Olympus abstract machine. Therefore, the Sieve of Eratosthenes\footnote{Due to the limited memory on the micro-core devices, the flag array size was reduced ($SIZE=4095$) on the Epiphany-III, MicroBlaze and PicoRV32 micro-cores, on the other CPUs, per the original benchmark, $SIZE=8190$.} (Sieve) benchmark was selected to determine the size efficiency and integer performance of Olympus relative to handwritten (native) C. Figure \ref{fig:sieveperformance} shows that, across compiler optimisation levels \texttt{-Os} and \texttt{-O3}, the Sieve benchmark displays a wider performance gap between the Olympus and native C kernels than was observed for the LINPACK benchmark, discussed in Section \ref{sec:codeperf}. The Olympus Sieve kernel performance ranges from approximately 1.4 times slower than native C on the Epiphany-III to over 5.5 times slower\footnote{Both bounds of the range at compiler optimisation level \texttt{-O3}.} on the RISCV64. For all CPUs apart from the AMD64, the difference between Olympus and native C kernel performance is smaller at compiler optimisation level \texttt{-O3} than at \texttt{-Os}. On the RISCV64, the native C Sieve kernel is 5.5 times faster than the Olympus kernel at \texttt{-Os} but is only 4 times faster at \texttt{-O3}.

 Whilst the kernel performance difference between the \texttt{-Os} and \texttt{-O3} GCC optimisation levels is greatest for the RISCV64, all of the RISC CPU Olympus kernels close the performance gap with the native C kernels at \texttt{-O3}. In comparison, the CISC AMD64 native C kernels are 1.7 times faster than Olympus at \texttt{-Os} and 2 times faster at \texttt{-O3}. This suggests that GCC is able to leverage the additional registers available on the RISCV64 over those available on the AMD64 to optimise the Olympus abstract machine code at \texttt{-O3} optimisation level. However, the results for the Epiphany-III, MIPS32 and SPARCv9 suggest that the additional registers available on the Epiphany-III do not provide an advantage over the 32 available on the the MIPS32 and SPARCv9. 


\ifdefined\DATATABLES
\begin{table}[h!]
\caption{Sieve benchmark native C and Olympus kernel runtime (seconds)}
\centering
\footnotesize
\begin{tabular}{|lllll|}
\hline
\textbf{CPU}  & \textbf{Olympus (-Os)} & \textbf{Olympus (-O3)} & \textbf{native C (-Os)}  & \textbf{native C (-O3)} \\ \hline
\textbf{Epiphany-III} & 4.972  & 3.475 & 2.760 & 1.996 \\
\textbf{MicroBlaze}  & 48.23 & 32.70 & 17.78 & 14.94  \\
\textbf{PicoRV32} & 358.64 & N/A\footnote{The Olympus Sieve kernel freezes at compiler optimisation level \texttt{-O3}.}
& 140.77 & 94.52 \\ 
\textbf{ARM32} & 12.53 & 9.135 & 4.629  & 3.443 \\
\textbf{MIPS32} & 13.79 & 9.211 & 3.683 & 2.858 \\
\textbf{AMD64}  & 3.895 & 2.868 & 2.319 & 1.409 \\
\textbf{SPARCv9} & 30.93 & 18.80 & 6.763 & 5.060 \\
\textbf{RISCV64} & 14.61 & 11.12 & 2.676 & 2.755 \\ \hline
\end{tabular}
\label{tbl:sieveperf}
\end{table}
\fi 

On the PicoRV32, the Olympus \texttt{-O3} kernel froze and did not return a value to the host, even though the kernel successfully executed when compiled at optimisation level \texttt{-Os}. As the LINPACK PicoRV32 kernels also failed to execute correctly at \texttt{-O3}, it is likely that the version of the RISC-V compiler used (riscv32-unknown-elf-gcc 8.2.0) is generating code that is invalid for the PicoRV32 at this level of optimisation.

\subsection{Sieve of Eratosthenes code size}\label{sec:sievesize}
\begin{figure}[h!]
	\centering
	\includegraphics[width=0.99\textwidth]{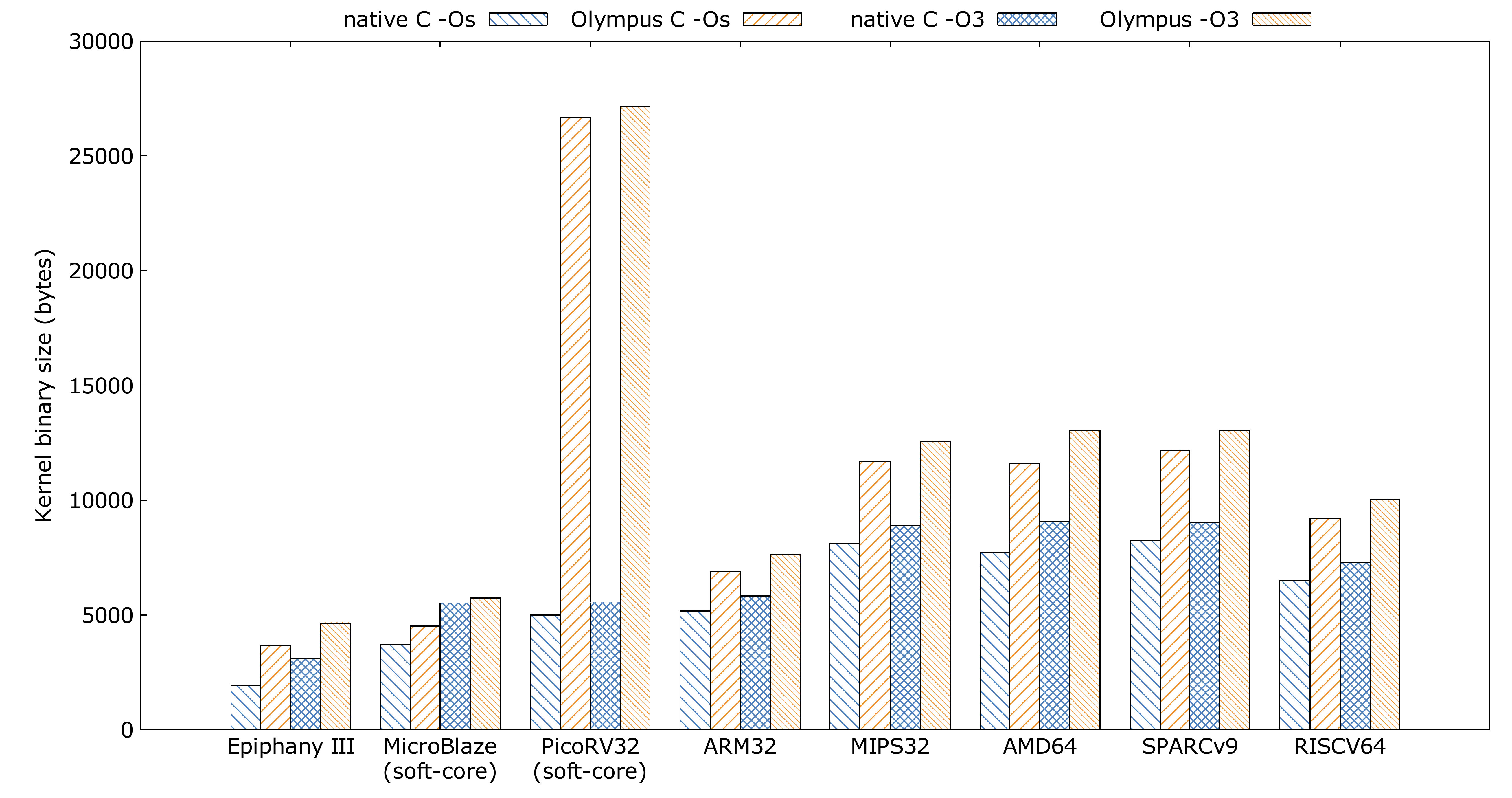}
	\caption{Sieve benchmark native C and Olympus kernel size}
	\label{fig:sievesize}
\end{figure}

\ifdefined\DATATABLES
\begin{table}[h!]
\caption{Sieve benchmark code size (bytes)}
\centering
\footnotesize
\begin{tabular}{|lllll|}
\hline
\textbf{CPU}  & \textbf{Olympus (-Os)} & \textbf{Olympus (-O3)} & \textbf{native C (-Os)}  & \textbf{native C (-O3)} \\ \hline
\textbf{Epiphany-III}  & 3678 & 4666 & 1938 & 3126 \\
\textbf{MicroBlaze}  & 4520 & 5724 & 3708  & 5516  \\
\textbf{PicoRV32} & 26668 & 27160 & 4980 & 5540 \\ 
\textbf{ARM32} & 6894  & 7602  & 5178  & 5826 \\
\textbf{MIPS32} & 11701  &  12582 & 8097  & 8894 \\
\textbf{AMD64}  & 11615  &  13039 & 7720 &  9080 \\
\textbf{SPARCv9} & 12195  &  13059 & 8231 &  9019 \\ 
\textbf{RISCV64} & 9185  & 10025 & 6470 & 7264 \\ \hline
\end{tabular}
\label{tbl:sievesize}
\end{table}
\fi 

Figure \ref{fig:sievesize} shows the size of the Sieve kernels compiled with \texttt{-Os} and \texttt{-O3} compiler optimisation levels for all CPUs. Whilst the Olympus kernel sizes are between near parity\footnote{MicroBlaze GCC optimisation level \texttt{-Os}.} and 1.9 times\footnote{Epiphany-III GCC optimisation level \texttt{-O3}.} that of the native C kernels for the other CPUs, the difference for the PicoRV32 is striking, with the Olympus kernel size around 5 times larger for both \texttt{-Os} and \texttt{-O3}. 
The Olympus kernel binary size for the PicoRV32 is explained by the fact that the GCC compiler allocates space 
in the kernel ELF file for the statically allocated C array used for the heap in the Olympus abstract machine. This is best illustrated by the size of the \texttt{.bss} segment reported by the GNU \texttt{size} utility for the Olympus Sieve kernel on the Epiphany-III, as shown in Listing \ref{lst:epiphanysize}, where the Olympus abstract machine heap is 24KB, and the 8MB default heap size of the RISCV64 desktop (threaded) kernel, as shown in Listing \ref{lst:rv64size}.

\begin{minipage}{0.95\linewidth}
\vspace{\baselineskip}
\begin{lstlisting}[style=cstyle, caption={Output of GNU \texttt{size} for Epiphany-III Olympus Sieve kernel}, label={lst:epiphanysize}]
   text    data     bss     dec     hex filename
   4666    1208   25336   31210    79ea e_task.elf
\end{lstlisting}
\end{minipage}

\begin{minipage}{0.95\linewidth}
\vspace{\baselineskip}
\begin{lstlisting}[style=cstyle, caption={Output of GNU \texttt{size} for RISCV64 Olympus Sieve kernel}, label={lst:rv64size}]
   text	   data	    bss	    dec	    hex	filename
   9185	    928	8001440	8011553	 7a3f21	threaded_sieve.elf
\end{lstlisting}
\end{minipage}

\begin{minipage}{0.95\linewidth}
\vspace{\baselineskip}
\begin{lstlisting}[style=cstyle, caption={Output of GNU \texttt{size} for PicoRV32 Olympus Sieve kernel}, label={lst:rv32size}]
   text    data     bss     dec     hex filename
  54668       0       0   54668    d58c rv_task.elf
\end{lstlisting}
\end{minipage}

In contrast, for the PicoRV32, as shown in Listing \ref{lst:rv32size}, there is is only a single \texttt{.text} segment, containing the executable code, static values, strings and the Olympus heap array. This is due to the custom GNU \emph{linker file} that is required to set up the memory map on the bare-metal PicoRV32 micro-core. The Epiphany-III and MicroBlaze micro-cores require similar custom linker files. However, the PicoRV32 file is unique in that the \texttt{KEEP} 
command is used to prevent the linker from performing \emph{dead code removal} on the \texttt{.text} segment, which is vital to ensure that the PicoRV32 register initialisation is performed. As the register initialisation subroutine is not referenced in the C source code, it would be removed by the GCC linker when the kernel binary is created, if the \texttt{KEEP} command was not used. As all functions are placed in the \texttt{.text} segment and no dead code removal is performed, all unused library functions will also be kept in the final binary, unlike the binaries for other CPUs. Although this is an issue for PicoRV32 binaries, it impacts both Olympus and native C kernels. Therefore, a more detailed discussion of possible mitigations for this issue will not be provided, except to highlight the benefits of the Olympus dynamic code loading mechanism discussed in \cite{jamieson_compact_2021}.

\subsection{Optimising loops}\label{sec:optimisingloops}
Whilst the performance of the Olympus abstract machine closes the gap with native C, the question remained as to whether the Olympus code generator could leverage the constrained vPython \texttt{for} loop to increase performance. Although it is considered \emph{unpythonic} to use \texttt{range} to provide an index variable to iterate through the elements of a list\cite{unpythonic_loop}, as shown in lines 2 and 3 of Listing \ref{lst:unpythonic}, rather than accessing an iterator directly as shown in lines 5 and 6, the iterator is \emph{immutable} and the list element is cannot be updated, whereas the unpythonic approach allows the list element to be updated. 

\begin{minipage}{0.95\linewidth}
\vspace{\baselineskip}
\begin{lstlisting}[style=pystyle, caption={\emph{Unpythonic} and \emph{Pythonic} list access}, label={lst:unpythonic}]
arr = [ "a", "b", "c"]
for i in range(0,len(arr)):
  arr[i] = "x"

for i in arr:
  i = "y"
\end{lstlisting}
\end{minipage}


Although a \texttt{while} loop with a manual index variable is often used in this case, the unpythonic \texttt{for} loop approach provides a performance benefit in vPython. As the iterator is managed by the Olympus abstract machine and not the programmer, the vPython \texttt{for} loop can leverage a native C local loop index variable, for example \texttt{\$iter\_i\$} in Listing \ref{lst:olyforloop}. This C variable not only controls the loop iteration but also is used to update the vPython list element, as shown in line 2 of Listing \ref{lst:olyforloop}. In contrast, the \texttt{while} loop requires a lookup of the index variable in the Olympus environment for both loop control and list element updates, as shown in lines 5, 6 and 7 of Listing \ref{lst:olyforloop}.

\begin{minipage}{0.95\linewidth}
\vspace{\baselineskip}
\begin{lstlisting}[style=cstyle, caption={Example Olympus abstract machine code for vPython loop constructs}, label={lst:olyforloop}]
FOR($iter_i$,0,LDI(ADDRL(2)),1)
STAI(ADDRL(4),$iter_i$,TRUE);
END

WHILE((LDI(ADDRL(10))<LDI(ADDRL(2))))
STAI(ADDRL(4),LDI(ADDRL(10)),TRUE);
STI(ADDRL(10),(LDI(ADDRL(10))+1));
END
\end{lstlisting}
\end{minipage}


Two vPython variants\footnote{Standalone versions, not run within the Eithne framework per Section \ref{sec:sievesperformance}.} of the Sieve benchmark were used to determine the performance benefits of the \texttt{for} loop over the \texttt{while} loop alternative. These were compiled at GCC optimisation levels \texttt{-Os} and \texttt{-O3}, and run on the RISCV64. A native C version of the Byte Sieve benchmark was also compiled at both optimisation levels and run for comparison with the vPython variants. As detailed in Table \ref{tbl:bytesieve}, the \texttt{for} loop variant of the vPython Byte Sieve benchmark is approximately 3 times faster at both \texttt{-Os} and \texttt{-O3} GCC optimisation levels than the \texttt{while} loop variant. Furthermore, the \texttt{for} loop variant closes the performance gap with native C to around 1.5 times slower from approximately 5 times slower for the \texttt{while} loop variant (both at GGC optimisation level \texttt{-Os}). 

\begin{table}[h!]
\caption{Byte Sieve benchmark runtime performance (seconds)}
\centering
\footnotesize
\begin{tabular}{|lll|}
\hline
\textbf{Code variant}  & \textbf{GCC -Os} & \textbf{GCC -O3} \\ \hline
\textbf{vPython \texttt{while}}  & 7.22 & 5.55 \\
\textbf{vPython \texttt{for}}  &  2.33  & 1.92  \\
\textbf{Native C} & 1.33  &  1.20 \\ \hline
\end{tabular}
\label{tbl:bytesieve}
\end{table}

The new version of the Olympus abstract machine for Vipera that separates the object addressing from operation within the mnemonics, not only enables direct access to native C variables, as shown in Listing \ref{lst:olyforloop}, to increase performance but also simplifies the implementation of object references within the abstract machine
, enabling the integration of Olympus applications with C frameworks, such as the Eithne benchmarking framework\cite{noauthor_maurice_nodate} and MPI (Message Passing Interface).

\ifdefined\RISCV
\subsection{Tailoring for RISC-V}
The Olympus abstract machine is cross-platform but can be easily tailored to support specific features of a CPU ISA, such as vector instructions. To illustrate the ease with which tailoring can be performed, we will discuss modifying the Olympus code generated for the vPython \texttt{for} loop to leverage the RISC-V temporary registers (\texttt{t0} - \texttt{t6}) for the loop \emph{iterator}. 
The register allocation is performed within the Olympus abstract machine code generation phase (Phase 4 in Figure \ref{fig:astflow}), when the generator for the \texttt{for} loops is entered, a register is \emph{popped} from the \emph{available register stack} and allocated to the loop iterator, with a corresponding \emph{push} of the register back onto the available stack performed when the loop is exited, allowing the register to be reused. The allocation algorithm is trivial; once all the RISC-V temporary registers are allocated, no attempt is made to \emph{spill} existing loop register allocations to enable the reuse of registers for a new \texttt{for} loop.
\fi 

\section{Conclusion}
Whilst the vPython virtual machine provided a productive environment to deploy parallel codes written in a dynamic language to micro-core architectures, the performance overhead of the interpreter limited its use for real-world codes. 
However, the Olympus abstract machine approach resulted in kernel performance that was comparable to or, in some cases could exceed, native C kernels, as confirmed for the LINPACK benchmark in Section \ref{sec:codeperf}, and, at a worst-case, was around five times slower than native C for the Sieve of Eratosthenes benchmark (Section \ref{sec:sievesperformance}). 
Crucially, as shown in Section \ref{sec:optimisingloops}, this gap can be lowered to just over 1.5 times slower by leveraging the \texttt{for} loop's native C iterator.
Furthermore, a single Python code is portable across these architectures, which is not the case for the standard C codes. 

Vipera has also addressed the portability of user codes and underlying runtime support. 
All of the benchmarks run unmodified across all the supported platforms and the Olympus abstract machine builds from a single codebase, which results in significant programmer productivity gains. All device-specific code is managed within the mnemonics and runtime support functions, with the generated Olympus abstract machine code remaining the same across all platforms. Furthermore, the vPython virtual machine was also shown to be portable to a number of micro-core architectures 
with the minimum of effort.

Further work includes exploring automatic memory management for data and code, optimisation of the Olympus abstract machine, automatic dynamic function selection for the dynamic loading support discussed in \cite{jamieson_compact_2021}, additional data types (byte arrays) to minimise the memory footprint of data and additional device support (GPUs and FPGAs) using OpenCL C
and Xilinx HLS C.

Whilst this paper has focused on the assessment of the Olympus code generation model using vPython, we also believe that Vipera has a wider applicability to other dynamic programming languages targeting micro-core architectures.





%
%

%
%
%
\bibliographystyle{splncs04}
\bibliography{xDSL.bib}
%
\end{document}